\documentclass[12pt]{article}

\usepackage{latexsym,amsmath,amssymb,theorem,epsfig,graphicx}

\usepackage{graphicx}
\usepackage{graphicx,subfigure}
\usepackage{epstopdf}
\usepackage{amssymb}
\usepackage{amsmath}
\usepackage{psfrag}
\usepackage{epsfig}
\usepackage{cancel}
 \allowdisplaybreaks[4]

\usepackage[all]{xy}

\DeclareFontFamily{OT1}{rsfs10}{}
\DeclareFontShape{OT1}{rsfs10}{m}{n}{ <-> rsfs10 }{}
\DeclareMathAlphabet{\mathscript}{OT1}{rsfs10}{m}{n}

\numberwithin{equation}{section}

\newcommand{\ns}{\normalsize}

\newcommand{\kk}{{\bf k}}

\newcommand{\comb}{(1-\epsilon-\epsilon_s)}

\newcommand{\es}{\epsilon_s}

\def\gsim{ \lower .75ex \hbox{$\sim$} \llap{\raise .27ex \hbox{$>$}} }
\def\lsim{ \lower .75ex \hbox{$\sim$} \llap{\raise .27ex \hbox{$<$}} }
\def\be{\begin{equation}}
\def\ee{\end{equation}}
\def\bea{\begin{eqnarray}}
\def\eea{\end{eqnarray}}

\usepackage{graphicx}

\newcommand{\ba}{\begin{array}}
\newcommand{\ea}{\end{array}}
\newcommand{\mn}{\mu\nu}
\newcommand{\ep}{\epsilon}
\begin{document}

\begin{titlepage}
\title{
  \hfill{\ns }  \\[1em]
   {\LARGE Bimetric structure formation: non-Gaussian predictions}
\\[1em] }
\author{Jo\~{a}o Magueijo$^{\rm 1}$, Johannes Noller$^{\rm 1}$ and Federico Piazza$^{\rm 2}$
     \\[0.5em]
     {\ns $^{\rm 1}$Theoretical Physics, Blackett Laboratory, Imperial College, London, SW7 2BZ, UK}\\
   {\ns $^{\rm 2}$Perimeter Institute for Theoretical Physics Waterloo, Ontario, N2L 2Y5, Canada}\\[0.3cm]}

\date{}

\maketitle

%\title{Bimetric structure formation: non-Gaussian predictions and
%beyond the minimal model}

%
%\newcommand{\addressImperial}{Theoretical Physics, Blackett Laboratory, Imperial College, London, SW7 2BZ, United Kingdom}

%\newcommand{\addressPI}{PI}

%\author{Jo\~{a}o Magueijo}
%%\email{magueijo@ic.ac.uk}
%\affiliation{\addressImperial}

%\author{Johannes Noller}
%\affiliation{\addressImperial}

%\author{Federico Piazza}
%\affiliation{\addressPI}

\begin{abstract}
The minimal bimetric theory employing a disformal transformation
between matter and gravity metrics is known to produce exactly
scale-invariant fluctuations. It has a purely equilateral
non-Gaussian signal, with an amplitude smaller than that of DBI
inflation (with opposite sign) but larger than standard inflation.
We consider non-minimal bimetric models, where the coupling $B$
appearing in the disformal transformation ${\hat g}_{\mn}= g_{\mn}
-B\partial_\mu\phi\partial_\nu\phi$ can run with $\phi$. For
power-law $B(\phi)$ these models predict tilted spectra. For each
value of the spectral index, a distinctive distortion to the
equilateral property can be found. The constraint between this
distortion and the spectral index can be seen as a ``consistency
relation'' for non-minimal bimetric models.
\end{abstract}

%\tightenlines
\thispagestyle{empty}

\end{titlepage}

\section{Introduction}
As more and more data pours into cosmology
(e.g.~\cite{reviewdata}) the pressure is on theorists and
model-builders to predict signatures that would unmistakably
falsify cosmological models. No longer is it good enough to face
the ``zeroth-order challenge'': that of producing scale-invariant
scalar fluctuations with the correct amplitude. Deviations from
exact scale-invariance will be detected without controversy (if
they do exist) in the near future. Fields tantalizingly beyond our
reach -- such as gravitational waves (tensor modes) or primordial
non-Gaussianity~\cite{kazuya} -- will hopefully become tangible over the next
decade. While inflation~\cite{infl} has dominated the theoretical
cosmology scene, interest has never floundered on alternatives,
such as pre-big bang cosmology~\cite{pbb,pbbpert}, ekpyrotic/cyclic models~\cite{ekp,multiekp} and a 
varying speed of light
(VSL)~\cite{vsl0,vsl1,vslreview}. In this paper we take recent
work on VSL~\cite{joao1,piao,joao2} one step beyond, investigating
departures from strict scale-invariance and non-Gaussian
signatures.

Inflationary mechanisms where a varying speed of sound $c_s$ plays a relevant role have already been explored~\cite{amend,kp}. 
In~\cite{kp}, it was shown that an adiabatic scale invariant spectrum is produced even if the expansion -- albeit still inflationary -- is far from exponential (the equation of state can be as far as $w \approx -3/4$ from de-Sitter), provided the speed of sound varies appropriately. A class of  contracting (``ekpyrotic") cosmologies where this mechanism can be applied was also found. Non-gaussianities were calculated in the limit of strict scale invariance ($n_s = 1$), they can be large in both the expanding and the contracting cases. The superluminal phase $c_s>1$ that we are considering here allows to consider expansions that are not even inflationary, as long as the condition $H^2 \propto c_s$ is satisfied (here $H$ is the Hubble parameter)~\cite{joao1}.

Perhaps the most elegant formulation of VSL is in the guise of
disformal bimetric theories, for which the speed of gravity
differs from the speed of light~\cite{bim,bim1}. In general this
is achieved by constructing the Einstein-Hilbert action from an
``Einstein'' metric $g_{\mn}$ (the Einstein frame), whilst
minimally coupling the matter fields to a ``matter'' metric $\hat
g_{\mn}$ (the matter frame), with: \be S=\frac{M_{Pl}^2}{2}\int
d^4 x {\sqrt {-g}}\, R[g_{\mu\nu}] + \int d^4 x {\sqrt {-{\hat
g}}}\, {\cal L}_m[{\hat g}_{\mu\nu},\Phi_{Matt}] +S_\phi \ee in
which $S_\phi$ determines the dynamics. The two metrics are
related by a non-conformal transformation, such as:
\begin{equation}\label{gmnhat}
{\hat g}_{\mu\nu}=g_{\mn}-B\partial_\mu\phi\partial_\nu\phi\; ,
\end{equation}
where $\phi$ is the ``bi-scalar'' field. Here $B$ is chosen to
have dimensions of $M^{-4}$, so that $\phi$ has dimensions of $M$.
(In this paper we use metrics with signature $-+ + +,$ and $B$ is
defined so that $B>0$ corresponds to a speed of light {\it larger}
than the speed of gravity.) In the most general case $B$ is an
arbitrary function of $\phi$, but in the minimal theory it's set
to a constant.

In these theories there are two light cones at any point, one for
massless matter particles,  another for gravitons. More generally
the two metrics may be seen as independent representations of the
local Lorentz group (or two non-equivalent tetrads~\cite{bim1}),
one valid for gravitons and the other for matter. Thus, different
Lorentz transformations must be used to transform among measurements
made with matter and gravity (or equivalently, with clocks and
rods operated by matter or gravitational phenomena). For this
reason causality paradoxes can be skirted~\cite{joao2,vikman}, in
contrast to straightforward tachyonic matter~\cite{kinney2}. This
argument makes the bimetric construction important in interpreting
superluminal structure formation models.

A number of dynamics $S_\phi$ for bimetric theories have been
considered. It was pointed out in~\cite{joao2} that a Klein-Gordon
equation for $\phi$ in the matter frame translates into DBI
dynamics in the gravity frame. Its corresponding Lagrangian,
however, is not the Klein-Gordon Lagrangian in the matter frame,
but simply a cosmological constant. (It was first noted
in~\cite{bim} that for bimetric theories a Klein-Gordon action in
the matter frame doesn't translate into a Klein-Gordon equation in
that frame). Thus, the simplest bi-scalar dynamics is generated by
\begin{equation}\label{slambda}
S_\phi=\int\sqrt{-\hat g}(-2\Lambda_m) \; ,
\end{equation}
and $\Lambda_m<0$ leads to a speed of light larger than the speed
of gravity. If we require the field $\phi$ to have Klein-Gordon
dynamics in the Einstein frame at low energies (when matter and
gravity frames coincide), we should consider additionally:
\begin{equation}\label{sphi}
S_\phi=\int\sqrt{-\hat g}\frac{1}{B} -\int\sqrt{-g}\frac{1}{B}
\end{equation}
i.e. a positive cosmological constant in the Einstein frame
balanced by a negative one in the matter frame, both with
magnitude tuned to $1/(2B)$. This action maps into the DBI action~\cite{eva,tong}
in the gravity frame with DBI coupling $f=-B$, as explained
in~\cite{joao2}. 
For a choice of sign where the speed of light in
the gravity frame is larger than one ($f=-B<0$), this is sometimes
labeled ``anti-DBI'' (although one should note that ``flipping'' the sign of $f$ means that this setup cannot be interpreted as portraying a relativistic probe brane embedded in a five dimensional bulk, as usual for DBI; for an earlier study of anti-DBI theories see also~\cite{vikman2}). Combined with a mass potential in the
gravity frame it leads to scaling solutions and scale-invariant
fluctuations~\cite{cusc,joao1}, without the need for accelerated
expansion or a contracting pre-Big-Bang phase. The investigation
of non-Gaussian features in models where $B$ is allowed to ''run''
is the main purpose of this paper.

In the presence of a speed of sound $c_s\neq 1$ for adiabatic perturbations, the three-point function contains terms proportional to the  power spectrum squared and terms which are further multiplied by a factor $c_s^{-2}$~\cite{seery,chen,kp}. 
In the sub-luminal case $c_s<1$, the ``$c_s^{-2}$" terms dominate. This is what enhances non-gaussianities in DBI inflation and makes the three-point function scale dependent~\cite{ladies,emiliano} in the case of a varying speed of sound~\cite{kp} (the combination $H^2/c_s$ is set to be constant by the scale invariance of the power spectrum and terms that appear with different powers of $c_s$ will therefore run with the scale).
In the opposite limit, the one of infinite speed of sound that we are considering here,  the ``$c_s^{-2}$" terms are suppressed and the remaining terms inherit the scale invariance from the power spectrum of the two point function.
The dimensionless quantity $f_{NL}$ is of order 1 and has opposite sign to DBI inflation, i.e. $f_{NL}\sim 1>0$ with the WMAP sign convention. 
While such a small signal is surely observationally challenging, our results are also appealing because of a consistency relation between the three- and two- point functions. In fact, in the $c_s\gg 1$ limit the three point function  [eq. \eqref{ouramplitude} below] becomes independent of the background  parameters (such as the equation of state $w$) and only mildly depends on the tilt $n_s-1$ of the power spectrum.

The structure of this paper is as follows. In Section~\ref{perts}
we review and extend results of cosmological perturbation theory
needed for the calculations in this paper. Then, in
Section~\ref{scinv} we explain how scale invariance may be
achieved in these models and derive the associated non-Gaussian
features. The non-minimal model is spelled out in
Section~\ref{tilt}, with the basic ``Gaussian'' predictions
presented as well as its non-Gaussian properties. Throughout the
paper we refer to two appendices, where we explain the more
technical aspects of the calculation. Finally in a concluding
section we examine our results from a wider perspective.

\section{Cosmological Perturbations}\label{perts}

Projecting (\ref{sphi}) onto the Einstein frame leads to the
(anti)-DBI action, which belongs to the general class of k-essence
models~\cite{kessence}. Cosmological perturbations have been
extensively studied for these models. Here we review the main
results, extending them wherever needed. The starting point is an
action of the form:
\begin{equation} \label{general}
S=\int {\rm d}^4x \sqrt{-g} \left[\frac{R}{2} +
P(X,\phi)\right]~,
\end{equation}
%where the scalar field Lagrangian is chosen of the DBI form,
%\begin{equation}
%P(X,\phi) = -f^{-1}(\phi)\sqrt{1-2f(\phi)X}+f^{-1}(\phi)-V(\phi)\, .
%\label{DBI}
%\ee
where  the pressure $P$ is a general function of the scalar field $\phi$ and the kinetic term $X=-\frac{1}{2}g^{\mu\nu}\partial_{\mu}\phi \partial_{\nu}\phi$.
The energy density reads
\begin{equation}
\rho = 2 X P_{,X} - P\,,
\end{equation}
while the speed of sound is given by
\begin{equation}
c_s^2 = \frac{P_{,X}}{\rho_{,X}}= \frac{P_{,X}}{P_{,X}+2X P_{,XX}}\,.
\end{equation}
In a FRW Universe of scale factor $a(t)$ and Hubble rate $H(t) = {\dot a}/a$ we define the slow-roll parameters as follows:
\begin{equation}
\epsilon \equiv - \frac{\dot H}{H^2},\quad
\es \equiv \frac{\dot c_s}{c_s H}, \quad
\eta \equiv \frac{\dot \epsilon}{\epsilon H}, \quad
\eta_s \equiv \frac{\dot \es}{\es H}.
\end{equation}
In the $n_s = 1$ scale-invariant case, non-Gaussianity has been calculated in~\cite{kp}. Here we generalize to the case of an arbitrary -- albeit small -- tilt and negligible running ($\eta \approx 0, \eta_s \approx 0$). The calculation of the three-point function also necessitates defining two further parameters derived
from $P(X,\phi)$~\cite{seery,chen}
\begin{eqnarray}
\Sigma&=&X P_{,X}+2X^2P_{,XX}  = \frac{H^2\epsilon}{c_s^2} ~,\\
\lambda&=& X^2P_{,XX}+\frac{2}{3}X^3P_{,XXX} ~.
\label{lambda}
\end{eqnarray}

At quadratic order, the action for the curvature perturbation $\zeta$ for general speed of sound models is given by~\cite{garrigamukhanov}
\be \label{actionzeta}
S_2 = \frac{M_{\rm Pl}^2}{2} \int {\rm d}^3x{\rm d}\tau \;z^2\left[ \left(\frac{d\zeta}{d\tau}\right)^2 - c_s^2(\vec{\nabla}\zeta)^2\right]\,,
\ee
where $\tau$ is conformal time, and $z$ is defined as usual by $z
= a \sqrt{2 \epsilon}/c_s$. In~\cite{kp} it was shown that it is
convenient to work in terms of the ``sound-horizon" time ${\rm d}y
= c_s {\rm d}\tau$ instead of $\tau$. Explicitly, when $\eta =
\eta_s = 0$, \be \label{y} y = \frac{c_s}{(\epsilon+\epsilon_s
-1)aH}\,. \ee It is useful to write the behavior in $y$-time of
some relevant quantities: \be a\sim (-y)^{\frac{1}{\epsilon_s +
\epsilon - 1}}\;;\qquad c_s \sim
(-y)^{\frac{\epsilon_s}{\epsilon_s + \epsilon - 1}}\;; \qquad H
\sim (-y)^{\frac{-\epsilon}{\epsilon_s + \epsilon - 1}}.
\label{const} \ee
The quadratic action then takes the form
\be
S = \frac{M_{\rm Pl}^2}{2}\int {\rm d}^3x{\rm d}y \;q^2\left[  \zeta'^2 -(\vec{\nabla}\zeta)^2\right]\,,
\ee
where $'\equiv {\rm d}/{\rm d}y$, and
\be
q \equiv \sqrt{c_s}z = \frac{a \sqrt{2 \epsilon}}{\sqrt{c_s}}\,.
\label{qdef}
\ee
Upon quantization the perturbations are expressed through creators and annihilators as follows,
\begin{equation} \label{modes}
\zeta(y, \kk) = u_k(y)a(\kk) + u_k^*(y) a^\dagger(-\kk) .
\end{equation}
However, in order to make the correct choice of vacuum, it is useful to refer to the canonically-normalized scalar variable $v = M_{\rm Pl} q\zeta$. Then the equations of motion for the Fourier modes
are given by
\be
v''_k +\left(k^2 - \frac{q''}{q}\right)v_k = 0\,.
\label{veqn}
\ee
It is well-known that this results in a scale-invariant spectrum if $q''/q = 2/y^2$. More generally, we have
\begin{equation}
\frac{q''}{q} = \frac{1}{y^2}\left(\nu^2 - \frac{1}{4}\right),
\end{equation}
and the solution for $v_k(y)$ corresponding to the Bunch Davis vacuum is
\begin{equation}
v_k(y) = \frac{\sqrt{\pi}}{2} \sqrt{- y} \, H^{(1)}_\nu (- k y),
\end{equation}
where $H^{(1)}_\nu$ are Hankel functions of the first kind. The relation between $n_s$, $\nu$ and $\epsilon$ and $\es$ is
\begin{equation}
n_s - 1 = 3 - 2 \nu = \frac{2 \epsilon + \es}{\es + \epsilon -1}\;
.
\end{equation}
Again, in the limit when $\nu = 3/2$,
\begin{equation}
v_k(y) = - \frac{1}{\sqrt{2 k}} \left(1 - \frac{i}{k y}\right) e^{- i k y} \qquad (\nu = 3/2)
\end{equation}
and  we recover the scale invariant spectrum.
Going back to the modes defined in \eqref{modes} we have
\begin{equation}
u_k(y) =  \frac{c_s^{1/2}}{a M_{\rm Pl} \sqrt{2 \epsilon}} v_k(y) =  \frac{c_s^{1/2}}{a M_{\rm Pl} 2^{3/2}} \sqrt{\frac{\pi}{\epsilon}} \sqrt{-y} H^{(1)}_\nu (- k y) .
\end{equation}
It is useful to adopt an approximate expression for the Hankel functions. By expanding at $|k y| \ll 1$ we obtain
\begin{equation} \label{Hankelexp}
H^{(1)}_\nu (- k y) = - i \, \frac{2^\nu \Gamma(\nu) (- k y)^{-\nu}}{\pi} [1+ i k y + {\cal O}(k y)^2] e^{-i k y}\,.
\end{equation}
which gives the following approximate expression for $u_k$,
\begin{equation} \label{u}
u_k(y) \approx - i \, \frac{H (\epsilon + \es -1)}{2 M_{\rm Pl} \sqrt{ c_s k^3 \epsilon}} \left(\frac{- k y}{2}\right)^{3/2 - \nu} (1 + i k y) e^{- i k y} .
\end{equation}
 In order to obtain~\eqref{u}, eq. ~\eqref{y} has been used, together with $\Gamma(\nu \approx 3/2) \approx \sqrt{\pi}/2$\footnote{In fact,  $\Gamma(\nu) = \sqrt{\pi}/2 \, [1+ 0.036 (\nu - 3/2) + \dots]$.}.
As expected, $u_k(y) \simeq$ const. in the $y\rightarrow 0$ limit. To check this explicitly we use~\eqref{const} and note that
\begin{equation} \label{andamento}
\frac{H}{c_s^{1/2}} \sim (-y)^{\nu - 3/2}\, .
\end{equation}
The derivative of $u_k(y)$ with respect to $y$ is also easily obtained:
\begin{equation}
u'_k(y) \approx -i \, \frac{H (\epsilon + \es -1)}{2 M_{\rm Pl} \sqrt{ c_s k^3 \epsilon}} \left(\frac{- k y}{2}\right)^{3/2 - \nu} k^2 y \ e^{- i k y} .
\end{equation}
Finally, the expression for the $\zeta$ Power Spectrum reads
\begin{equation}
P_\zeta \equiv \frac{1}{2\pi^2}k^3\left\vert\zeta_k\right\vert^2 = \frac{(\epsilon_s + \epsilon-1)^2 \, 2^{2 \nu - 3}}{2 (2 \pi)^2\epsilon}\frac{{\bar H}^2}{{\bar c}_sM_{\rm Pl}^2}\,,
\label{zetaPk}
\end{equation}
where the bar symbol means that the corresponding quantity has to be evaluated, for each mode $k$, at sound horizon exit, \emph{i.e.}, when $y = k^{-1}$.

\section{The Scale Invariant limit}\label{scinv}

For the bimetric theories discussed in the introduction, in the
Einstein frame the action takes the form: \be P(X,\phi) =
-f^{-1}(\phi)\sqrt{1-2f(\phi)X}+f^{-1}(\phi)-V(\phi)\, \label{DBI}
\ee where $X=-\frac{1}{2}\partial_\mu\phi\partial^\mu\phi$. 
Scaling solutions of this action have been studied
in~\cite{kinney1,kp,kinney2}. 
In particular, a scale invariant
spectrum of primordial perturbations is produced if
\begin{eqnarray}
f(\phi) &=& -\frac{3}{V_0} \left[ \frac{1}{4} - \frac{1}{\epsilon^2}\left(\frac{\phi}{M_P}\right)^{-4}\right] \\
V(\phi) &=& V_0 \left(\frac{\phi}{M_{\rm Pl}}\right)^{2}\left(1- \frac{2\epsilon}{3}\frac{1}{1+ \frac{\epsilon_s^2\phi^2}{8\epsilon M_{\rm Pl}^2}}\right) \nonumber\\
&&\simeq V_0 \left[\left(\frac{\phi}{M_P}\right)^2 - \frac{4}{3} + \dots\right]\, ,
\end{eqnarray}
where we have made use of the slow-roll parameter $\ep =
-\dot{H}/H^2$. In the large $\phi$ limit (which also corresponds
to the large $c_s$ limit) we recover the form discussed
in~\cite{joao2}: $f(\phi) \simeq - B<0$, $V(\phi) \sim \phi^2$.
This corresponds in general to $c_s\propto \rho$. In such a strict
$c_s\rightarrow\infty$ limit the amplitude of the three point
function ${\cal A}$ can be read straightforwardly
from~\eqref{3ampI} (we refer the reader to Appendices A
and B for the more gruesome technical details). Comparing with the
cubic effective action \eqref{action3} we find that only the
${\cal A}_{\zeta \dot\zeta^2}$ and ${\cal A}_{\zeta(\partial
\zeta)^2}$ terms are not subdominant as $c_s\rightarrow\infty$.
The resulting total amplitude is independent of the parameters
($w$ or $\epsilon$) and reads
\begin{equation}
{\cal A}_{c_s\rightarrow \infty} = - \frac{1}{8}\sum_i k_i^3 + \frac{1}{K} \sum_{i<j} k_i^2 k_j^2 -\frac{1}{2 K^2} \sum_{i\neq j}k_i^2 k_j^3\, . \label{SIamp}
\end{equation}
This is precisely the equilateral shape, peaking for $k_1 = k_2 = k_3$, that is also obtained in the scaling solutions considered in~\cite{kp} in the  $\epsilon \to 0$, $\alpha \to 0$ limit. More specifically one obtains
\begin{equation}
{\cal A}_{\ep \to 0} = {\left(1 - \frac{1}{c_s^2}\right)} {\cal
A}_{c_s \to \infty}\ +\ {\cal O}(n_s -1)\; , \ee%
This shows that in the minimal bimetric model the dimensionless quantity $f_{NL}$ is of order
1 and has opposite sign to DBI inflation, i.e. $f_{NL}\sim 1>0$.
Thus the model is quite distinct in this respect to standard
inflation (for which $f_{NL}\sim\ep\sim 0.1$) and DBI inflation
(for which $f_{NL}\sim -100$ is a distinct possibility.) Notice
there's been some confusion~\cite{tong}, both among theorists and
observers, regarding the sign of $f_{NL}$. Here we adopt the
convention used by WMAP, where positive $f_{\rm NL}$ physically corresponds to negative-skewness for the temperature fluctuations:  we assign a negative $f_{NL}$ to DBI
inflation, so that $f_{NL}>0$ for the anti-DBI models under
consideration.

\section{Beyond the minimal model}\label{tilt}

It could be that the parameter $B$ appearing in the disformal
transformation (\ref{gmnhat}) is itself a function of $\phi$. In
this Section we show that non-minimal theories with power-law
$B(\phi)$ lead to tilted spectra, without running. Naturally, more
complicated $B(\phi)$ would lead to more complex spectra, so one
can't say that absence of running is a general feature of these
models.

First we note that the speed of sound in the gravity frame is:
\begin{equation}
c^2_s=\frac{K_{,X}}{K_{,X}+2XK_{,XX}}=1+2B X
\ee
whereas the density and pressure are:
\bea
\rho&=&2XK_X-p=\frac{1}{B}{\left(1-\frac{1}{c_s}\right)} +V\\
p&=& K-V=\frac{1}{B}(c_s-1) -V
\eea
Thus scaling solutions may be obtained for a variety of potentials,
with the property that
\bea
\epsilon&=&-\frac{\dot H}{H^2}=\frac{3}{2}(1+w)\\
\epsilon_s&=&\frac{\dot c_s}{c_s H} \eea are constant. As
explained in the Appendix A of reference~\cite{kp} the constancy
of $\epsilon$ and $\epsilon_s$ permits a simple integration into
$c_s=c_s(\phi)$ and $H=H(\phi)$. The Friedmann equations then give
a solution for $V(\phi)$ and $f(\phi)$. 
\bea V&=& V_0\left(\frac{\phi}{M_{\rm Pl}}\right)^{-4\epsilon/\epsilon_s}\left(1- \frac{2\epsilon}{3}\frac{1}{1+ \frac{\epsilon_s^2\phi^2}{8\epsilon M_{\rm Pl}^2}}\right) \\
& = & 
V_0{\left(\frac{\phi}{M_{\rm Pl}}\right)}^{-4\epsilon / \epsilon_s} \left[ 1 - \frac{16 \epsilon^2 M_{\rm Pl}^2}{3 \epsilon_s^2 \phi^2} + {\cal O} \left(\frac{\phi}{M_{\rm Pl}}\right)^{-2}\right] \\[2.5mm]
f(\phi) &=& \frac{12}{V_0\epsilon_s^2}\left(\frac{\phi}{M_{\rm Pl}}\right)^{\frac{4\epsilon}{\epsilon_s}-2}\left(1-\frac{\epsilon_s^4\phi^4}{64\epsilon^2M_{\rm Pl}^4}\right) \\
&=& - \frac{3\epsilon_s^2}{16\epsilon^2 V_0}
{\left(\frac{\phi}{M_{\rm Pl}}\right)}^{2+\frac{4\epsilon}{\epsilon_s}} \left[1+{\cal O}\left( \frac{\phi}{M_{\rm Pl}}\right)^{-4}\right]
\eea
Using $B=-f>0$ and in the limit $c_s\gg 1$ we therefore obtain 
\bea
V&=&V_0{\left(\frac{\phi}{M_{Pl}}\right)}^{-\frac{4\epsilon}{\epsilon_s}}\\
B&=&-f=\frac{3\epsilon_s^2 M_{Pl}^4}{16\epsilon^2 V_0}
{\left(\frac{\phi}{M_{Pl}}\right)}^{2+\frac{4\epsilon}{\epsilon_s}}
\eea
Following the calculation in~\cite{joao1} we find that for these
solutions the spectral index is:
\begin{equation}
n_S-1=\frac{\ep_s+2\ep}{\ep_s+\ep-1}\; . \ee Whilst
scale-invariance is associated with the universal law \be
c_s\propto \rho \ee and also $f=1$ and a quadratic potential {\it for
all equations of state}, the same doesn't happen if we depart from
scale-invariance. Indeed exact scale invariance requires
$\epsilon_s=2\ep$ so that these parameters fall out of conditions
for the spectral index; but this doesn't happen as soon as
$n_s\neq 1$. Note, however,  that {\it any} scaling solution has:
\be\label{csphi}
c_s(\phi)=\frac{\ep_s^2}{8\ep}\phi^2
\ee
an expression that will be essential in evaluating
non-Gaussianities.

%\section{Non-Gaussianity in the non-minimal model}\label{ngtilt}

In Appendix B we present expressions for the Non-Gaussian
amplitude ${\cal A}$ for general $c_s$ profiles. Our calculations there consequently apply to both subluminal and superluminal cases, as long as scaling solutions with constant $\epsilon$ and $\epsilon_s$ are considered. Once a solution for $c_s$ is provided, these uniquely determine the Non-Gaussian signature.
However, here we exclusively focus on the $c_s \to \infty$ case relevant in
the bimetric context. Considering further the relation
(\ref{csphi}), fixing $c_s(\phi)$,
%\be
%c_s(\phi)=\frac{\ep_s^2}{8\ep}\phi^2,
%\ee
we have that in the large $\phi$ limit the appropriate
Non-Gaussian amplitude to compute is still ${\cal A}_{\bar{c}_s
\to \infty}$. 

\begin{figure}[t]
  \begin{center}
     \subfigure[$-{\cal A}(1,x_2,x_3)/(x_2 x_3)$ for $n_s = 1$]{\label{fig2-a}\includegraphics[width=3in]{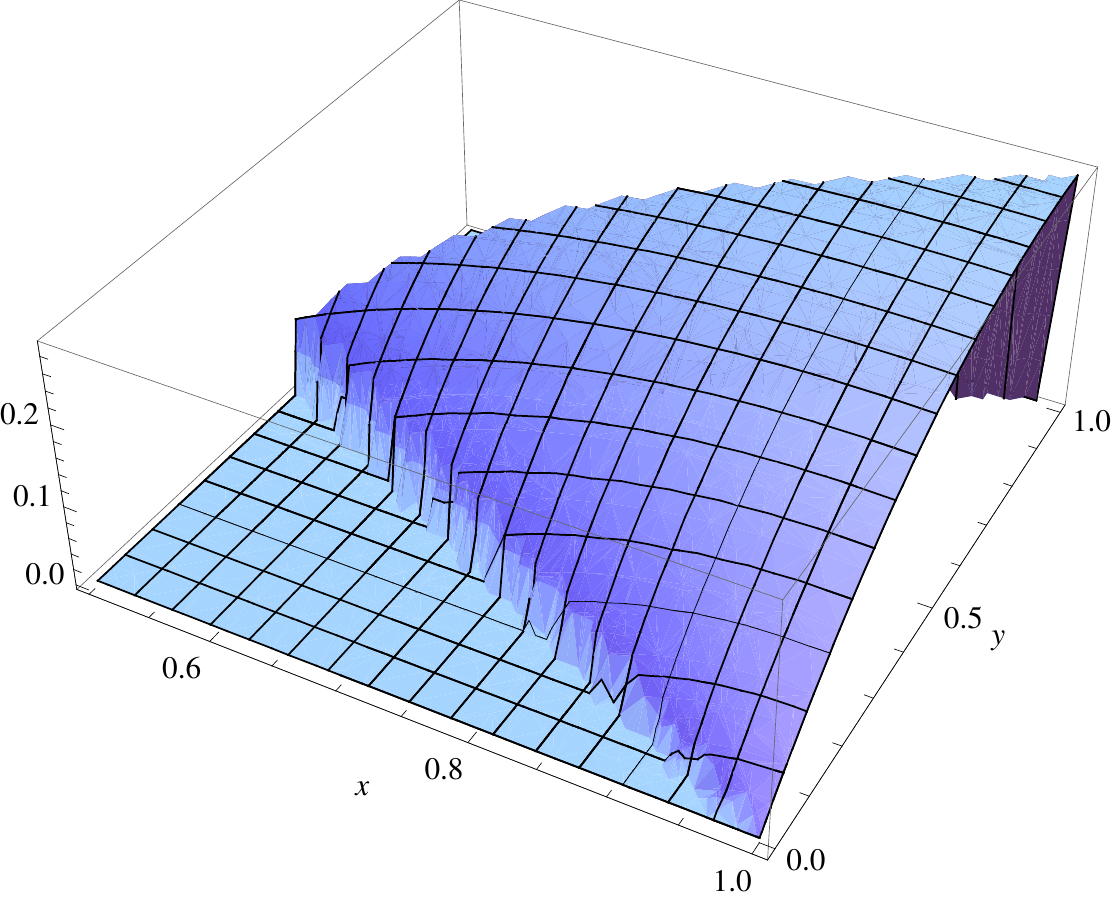}}
    \subfigure[$-{\cal A}(1,x_2,x_3)/(x_2 x_3)$ for $n_s = 0.96$]{\label{fig2-b}\includegraphics[width=3in]{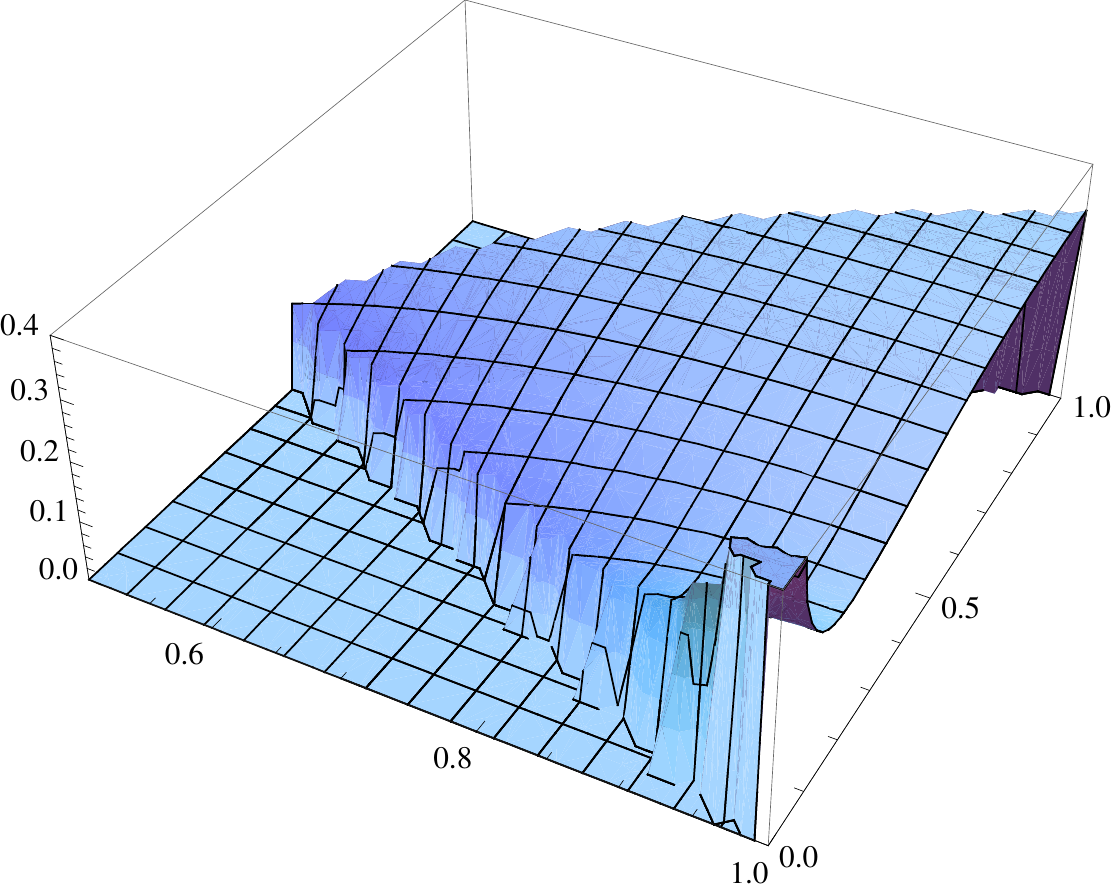}} \\
      \end{center}
\caption{We plot the non-Gaussian amplitude from Eq.~\eqref{ouramplitude} $-{\cal A}(1,x_2,x_3)/(x_2 x_3)$ for $n_s = 1$ (left) and $n_s = 0.96$  (right).}
   \label{fig2}
\end{figure}

Combining terms from the general results of Appendix \ref{app:2}, eqs. \eqref{3ampI} and \eqref{amplitudes} and taking the small tilt ($n_s - 1 \ll 1$) and $c_s \to \infty$ limits, we find  
\begin{align}
{\cal A} \ = &\  \left(\frac{k_1 k_2 k_3}{2 K^3}\right)^{n_s -1} \left[- \frac{1}{8}\sum_i k_i^3 + \frac{1}{K} \sum_{i<j} k_i^2 k_j^2 -\frac{1}{2 K^2} \sum_{i\neq j}k_i^2 k_j^3\, \right. \nonumber \\
&\ + (n_s -1)\left(- \frac{1}{8}\sum_i k_i^3 - \frac{1}{8} \sum_{i\neq j} k_i k_j^2 + \frac{1}{8}k_1 k_2 k_3  + \frac{1}{2 K} \sum_{i<j} k_i^2 k_j^2- \frac{1}{2 K^2} \sum_{i\neq j}k_i^2 k_j^3\right)\nonumber \\ 
&\ \left. + {\cal O}\left(\frac{1}{c_s^2}\right) \right] \; , \label{ouramplitude}
\end{align}
where the only dependence on $\epsilon$ and $\epsilon_s$ appears either in the ``observable" combination $n_s -1$ or in the subleading ${\cal O}(1/c_s^2)$ terms.  Upon approaching scale
invariance only the first line inside the square brackets stays relevant, ensuring that ${\cal A}$ reduces to the equilateral amplitude \eqref{SIamp} as required.

The amplitude~\eqref{ouramplitude} is plotted in Figure 1 and peaks in the equilateral limit $k_1 = k_2 = k_3$. In the local limit $k_1 \ll k_2, k_3$, on the other hand, the first line inside the square brackets of Eq. 
\eqref{ouramplitude} goes to zero. In agreement with the consistency relation~\cite{malda,consistency} we then have
\begin{equation}
{\cal A}_{k_1 \ll k_2, k_3} \ \approx \ -\frac{1}{2} (n_s -1) \left(\frac{k_1}{k_2}\right)^{n_s -1}
\end{equation}

The predictive power of our result lies in establishing a consistency relationship between $n_s$ and ${\cal
A}$. In fact, we find a distinctive Non-Gaussian signal for any given
spectral index $n_s$. Whilst the overall non-Gaussian amplitude ${\cal A}$ still peaks in the equilateral limit $k_1 = k_2 = k_3$ in both red- and blue-tilted cases, its shape is modified when compared with the scale invariant limit.
Illustrating this point, Figure 1a shows ${\cal A}$ for the exactly scale invariant case $n_s = 1$, whereas in Figure 1b we plot ${\cal A}$ for a red-tilted power spectrum with $n_s = 0.96$. 
Specifically we find an ${\cal A}(n_s = 0.96) = {\cal A}(n_s = 1) + {\Delta \cal A}$, where $\Delta \cal A$ is approximately one order of magnitude smaller than ${\cal A}(n_s = 1)$.

\section{Conclusions}\label{concs}
Strict scale-invariance has been associated with superluminal
bimetric models, where the speed of light is larger than the speed
of gravity in the early Universe~\cite{joao1,joao2}. Indeed this
is a feature of the minimal bimetric model, but in this paper we
showed how tilted spectra, red or blue, could be generated by a
non-minimal bi-scalar coupling $B(\phi)$. At first this might
suggest we've fallen into the ``theory of anything'' trap, but
it's not the case. 
A unique non-Gaussian shape is predicted for any value of the spectral index, with distinct distortions away from the scale-invariant equilateral shape appearing for each of the tilted cases. These distortions can be seen as ``consistency conditions'' for this class of models.
This is particularly relevant given the absence
of gravitational waves for all bimetric models of this kind. (Note
that these models solve the horizon problem for matter but not for
gravity, so tensor modes don't start their lives inside the
horizon.)

One might wonder where the proposed running coupling $B(\phi)$
comes from. First note that we don't need the full (anti-)DBI 
action (\ref{DBI}) resulting from (\ref{sphi}), unless we impose
Klein-Gordon dynamics for $\phi$ in the Einstein frame at low
energies. This may not be necessary, and if we relax this
requirement all we need is (\ref{slambda}), i.e. a negative
cosmological constant $\Lambda_m$ in the matter frame (which, we
stress, does {\it not} lead to an AdS solution). In fact, if we
relax the low-energy requirement, $\Lambda_m$ doesn't even need to
be related to $B$. If, however, we do insist on Klein-Gordon
dynamics for $\phi$ in the Einstein frame at low energies, then
the negative matter frame cosmological constant should be exactly
balanced by a positive Einstein frame cosmological constant, and
their common magnitude should be $1/(2B)$.

A number of interesting theoretical connections can be made. In
the context of emergent geometry, it's been pointed out that
different emergent metrics may apply to bosons and
fermions~\cite{fotini}. The fact that the vacuum energy is
negative for fermions and positive for bosons suggests an action
of the proposed form, with a speed of light larger than the speed
of gravity (i.e. an anti-DBI action in the Einstein frame). Also
these models become asymptotically a cuscaton~\cite{cusc} model, a
feature that may be used to support the view that they are a
UV-complete alternative to inflation.  Finally, it is possible
that this construction results from an entirely different set up,
such as deformed dispersion relations~\cite{csdsr}. It is
interesting that the dispersion relations needed for
scale-invariance are of the same form as those discussed in the
context of Horava-Lifshitz theory~\cite{horava}. More generally a
connection with deformed special relativity remains to be fully
explored~\cite{dsr,ljprl}. Absence of exact scale-invariance could
then be a major clue into the foundations of these theories.

While work on these theoretical ramifications is an interesting
motivation, and should be pursued further, in this paper we
focused on the phenomenology of these models. Measuring the shape
of the three-point correlator (as opposed to a quantity as muddled
as $f_{NL}$) poses an interesting observational challenge. The
fact that the matter appears coupled to the measurement of $n_S$
makes these models an interesting target for future experimental
work.

\section{Acknowledgments}
We'd like to thank N. Afshordi, J. Khoury and F. Vernizzi for helpful discussions. JM thanks
the Perimeter Institute for hospitality. JN is supported by an STFC studentship. FP is supported in part by the Government of Canada through NSERC and by the Province of Ontario through the Ministry of Research \& Innovation. 

\appendix
\section{Appendix: The cubic action} \label{A1}

It is useful to report here the cubic effective action derived in~\cite{seery,chen}.
The result is valid outside of the slow-roll approximation and for any time-dependent sound speed:
\begin{eqnarray} \label{action3}
S_3&=&{M_{Pl}}^2\int {\rm d}t {\rm d}^3x \left\{
-a^3 \left[\Sigma\left(1-\frac{1}{c_s^2}\right)+2\lambda\right] \frac{\dot{\zeta}^3}{H^3}
+\frac{a^3\epsilon}{c_s^4}(\epsilon-3+3c_s^2)\zeta\dot{\zeta}^2 \right.
\nonumber \\ &+&
\frac{a\epsilon}{c_s^2}(\epsilon-2\es+1-c_s^2)\zeta(\partial\zeta)^2-
2a \frac{\epsilon}{c_s^2}\dot{\zeta}(\partial
\zeta)(\partial \chi) \nonumber \\ &+& \left.
\frac{a^3\epsilon}{2c_s^2}\frac{d}{dt}\left(\frac{\eta}{c_s^2}\right)\zeta^2\dot{\zeta}
+\frac{\epsilon}{2a}(\partial\zeta)(\partial
\chi) \partial^2 \chi +\frac{\epsilon}{4a}(\partial^2\zeta)(\partial
\chi)^2+ 2 f(\zeta)\left.\frac{\delta L}{\delta \zeta}\right\vert_1 \right\} ~,
\end{eqnarray}
where dots denote derivatives with respect to proper time $t$, $\partial$ is a spatial derivative,
and $\chi$ is defined as
\begin{equation}
\partial^2 \chi = \frac{a^2 \epsilon}{c_s^2}\dot{\zeta}\,.
\end{equation}
Meanwhile, in the last term $\frac{\delta L}{\delta\zeta}|_1$ denotes the variation of the
quadratic action with respect to the perturbation $\zeta$:
\begin{eqnarray}
\left.\frac{\delta
L}{\delta\zeta}\right\vert_1 &=& a
\left( \frac{d\partial^2\chi}{dt}+H\partial^2\chi
-\epsilon\partial^2\zeta \right) ~,
\end{eqnarray}
\begin{eqnarray} \label{redefinition}
f(\zeta)&=&\frac{\eta}{4c_s^2}\zeta^2+\frac{1}{c_s^2H}\zeta\dot{\zeta}+
\frac{1}{4a^2H^2}[-(\partial\zeta)(\partial\zeta)+\partial^{-2}(\partial_i\partial_j(\partial_i\zeta\partial_j\zeta))] \nonumber \\
&+&
\frac{1}{2a^2H}[(\partial\zeta)(\partial\chi)-\partial^{-2}(\partial_i\partial_j(\partial_i\zeta\partial_j\chi))] ~,
\end{eqnarray}
where $\partial^{-2}$ is the inverse Laplacian. Since $\frac{\delta L}{\delta\zeta}|_1$ is proportional to
the linearized equations of motion, it can be absorbed by a field redefinition
\begin{eqnarray}
\zeta \rightarrow \zeta_n+f(\zeta_n) ~.
\end{eqnarray}

\section{Appendix: The three-point function} \label{app:2}

The three point function can be calculated by following the same method of~\cite{kp} and generalizing it.
The standard calculation~\cite{malda,seery,chen}, at first order in perturbation theory and in the interaction picture, leads to
\begin{equation} \label{interaction}
\langle
\zeta(t,\textbf{k}_1)\zeta(t,\textbf{k}_2)\zeta(t,\textbf{k}_3)\rangle=
-i\int_{t_0}^{t}{\rm d}t^{\prime}\langle[
\zeta(t,\textbf{k}_1)\zeta(t,\textbf{k}_2)\zeta(t,\textbf{k}_3),H_{\rm int}(t^{\prime})]\rangle ~,
\end{equation}
where $H_{\rm int}$ is the Hamiltonian evaluated at third order in the perturbations and is directly derivable from~\eqref{action3} and vacuum expectation values are evaluated w.r.t. the interacting vacuum $|\Omega \rangle$. By using~\eqref{modes} and applying the commutation relations
$[a(\kk), a^\dagger(\kk')] = (2 \pi)^3 \delta^3(\kk - \kk')$,
we can calculate the three point function for each term appearing in the action~\eqref{action3}. It is useful to follow in detail the calculation for the ``$\zeta \dot\zeta^2$" piece. We have:
\begin{multline}
\langle \zeta(\textbf{k}_1)\zeta(\textbf{k}_2)\zeta(\textbf{k}_3)\rangle_{\zeta \dot\zeta^2}\, =\,
i (2 \pi)^3 \delta^3(\kk_1+\kk_2+\kk_3) u_{k_1}(y_{\rm end})u_{k_2}(y_{\rm end})u_{k_3}(y_{\rm end}) \\ \times
\int_{-\infty+i\varepsilon}^{y_{\rm end}} {\rm d} y \frac{c_s}{a} \frac{a^3 \epsilon}{c_s^4} (\epsilon - 3 + 3 c_s^2)
u_{k_1}^*(y) \frac{d u_{k_2}^*(y)}{d y}\frac{d u_{k_3}^*(y)}{d y} + {\rm perm.} + {\rm c.c.}
\end{multline}
The subscript ``end" means that the quantity has to be evaluated at the end of ``inflation". We now substitute~\eqref{u} and use ~\eqref{andamento} to take some time-independent combinations outside the integral and evaluate them at $y = y_{\rm end}$:
\begin{multline}
\langle \zeta(\textbf{k}_1)\zeta(\textbf{k}_2)\zeta(\textbf{k}_3)\rangle_{\zeta \dot\zeta^2}\, =\, i (2 \pi)^3 \delta^3(\kk_1+\kk_2+\kk_3)
\frac{H_{\rm end}^6 \comb^6 2^{6\nu - 9}}{4^3 {M_{Pl}}^{4}\, \epsilon^{2}\, c_{s\, \rm end}^{\, 3}}\frac{(k_1 k_2 k_3)^{3 -2\nu}}{\Pi_j k_j^3} |y_{\rm end}|^{6\left(\frac{3}{2} - \nu\right)} \\
\times \int_{-\infty+i\varepsilon}^{{y_{\rm end}}} {\rm d}y (\epsilon - 3 + 3 c_s^2)
\frac{a^2}{{c}_s^{3}} (1 - i k_1 y) k_2^2 k_3^2 y^2 e^{i  K y}  + {\rm perm.} + {\rm c.c.}\,,
\end{multline}
where we have dropped a factor of $\Pi_j (1 + i k_j y_{end})e^{-i K y_{end}}$ as this will be negligibly small in the limit $|ky| << 1$ where the truncated Hankel function expansion \eqref{Hankelexp} is valid.

By using~\eqref{const} we can finally express the time dependent quantities inside the integrals as power-laws in $y$. It is useful to report some of the basic results of~\cite{kp} for integrals of this type. By calling
\begin{equation} \label{integral}
{\cal C} = \int_{-\infty + i\varepsilon}^{y_{\rm end}} {\rm d} y \left(\frac{y}{y_{\rm end}}\right)^\gamma (- i y)^n e^{i K y}\, .
\end{equation}
For $\gamma + n >-2$ the imaginary part of \eqref{integral} is convergent as $y_{\rm end}\rightarrow 0$. In this case we can approximately extend the upper limit of integration to $0$, which amounts to neglecting terms of higher order in $(k |y_{\rm end}|)$. We thus obtain
\begin{equation} \label{Cconvergent}
{\rm Im}\, {\cal C} = - (K |y_{\rm end}|)^{-\gamma} \cos\frac{\gamma \pi}{2} \Gamma(1+ \gamma + n) K^{-n-1}\, .
\end{equation}
The two types of behavior that we encounter are, in particular,
\begin{eqnarray}
\frac{a^2 y^2}{{c}_s} &=& \frac{c_s}{\comb^2 H^2}  =  \frac{c_{s\, \rm end}}{\comb^2 H_{\rm end}^2}  \left(\frac{y}{{y_{\rm end}}}\right)^{\alpha_1}\\
\frac{a^2 y^2}{{c}_s^{3}} &=& \frac{1}{\comb^2 H^2 c_s}  =  \frac{1}{\comb^2 H_{\rm end}^2 c_{s\, \rm end}}   \left(\frac{y}{{y_{\rm end}}}\right)^{\alpha_2},
\end{eqnarray}
where
\begin{eqnarray}
\alpha_1 & = & n_s - 1= 3 - 2\nu = \frac{2 \epsilon + \es}{\es + \epsilon -1}\\
\alpha_2 &  = & \frac{2 \epsilon - \es}{\es + \epsilon -1}
\end{eqnarray}

By using the above formulas and re-expressing everything in terms of quantities calculated at sound horizon crossing (\emph{i.e.} when, by convention, $y = K^{-1}$) we finally obtain
\begin{multline}
\langle \zeta(\textbf{k}_1)\zeta(\textbf{k}_2)\zeta(\textbf{k}_3)\rangle_{\zeta \dot\zeta^2}\, =\, (2 \pi)^3 \delta^3(\kk_1+\kk_2+\kk_3)
\frac{{\bar H}^4 (\epsilon + \es -1)^4 2^{6 \nu - 9}}{16 {M_{Pl}}^4 \epsilon^{2}\, {\bar c}_s^{4}}\frac{1}{\Pi_j k_j^3} \frac{(\Pi_j k_j^3)^{3-2\nu}}{K^{9 - 6 \nu}} \\
\times \frac{k_2^2 k_3^2}{K}
\left\{(\epsilon-3) \cos\frac{\alpha_2 \pi}{2} \Gamma(1+\alpha_2)\left[1+ (1+ \alpha_2) \frac{k_1}{K}\right] + 3 {\bar c}_s^2 \cos\frac{\alpha_1 \pi}{2} \Gamma(1+\alpha_1)\left[1+ (1+\alpha_1)\frac{k_1}{K}\right] \right\}+{\rm sym}.
\end{multline}

The three point function is conveniently expressed,
after factoring out appropriate powers of the power spectrum, through the amplitude ${\cal A}$,
\begin{equation} \label{amplidef}
\langle \zeta(\textbf{k}_1)\zeta(\textbf{k}_2)\zeta(\textbf{k}_3)\rangle = (2\pi)^7
\delta^3(\kk_1+\kk_2+\kk_3) P_\zeta^{\;2} \frac{1}{\Pi_j k_j^3}{\cal A}\,.
\end{equation}
Again, by convention, the power spectrum $P_\zeta$ in the above formula is calculated for the mode $K$.
By following the same steps above for each of the terms in the action~\eqref{action3} we obtain

\begin{align} %\label{decomposedAmp}
{\cal A}_{\zeta \dot\zeta^2}\, =&\ 
\frac{1}{4 {\bar c}_s^{2}} \left(\frac{k_1k_2k_3}{2 K^3}\right)^{3-2\nu} \left[(\epsilon - 3) {\cal I}_{\zeta \dot\zeta^2}(\alpha_2) + 3 {\bar c}_s^2 {\cal I}_{\zeta \dot\zeta^2}(\alpha_1)   \right] \nonumber\;; \\[2.5mm] \nonumber
{\cal A}_{\zeta(\partial \zeta)^2} =& \
\frac{1}{8 {\bar c}_s^{2}} \left(\frac{k_1k_2k_3}{2 K^3}\right)^{3-2\nu}  \left[(\epsilon - 2 \es +1) {\cal I}_{\zeta(\partial \zeta)^2}(\alpha_2) - {\bar c}_s^2 {\cal I}_{\zeta(\partial \zeta)^2}(\alpha_1)   \right] \nonumber\;; \\[2.5mm] \nonumber
{\cal A}_{\dot \zeta \partial \zeta \partial \chi} =& \ \nonumber
\frac{  1}{4 {\bar c}_s^{2}} \left(\frac{k_1k_2k_3}{2 K^3}\right)^{3-2\nu}  \left[ - \epsilon \, {\cal I}_{\dot \zeta \partial \zeta \partial \chi}(\alpha_2) \right] \;; \\[2.5mm]
{\cal A}_{\epsilon^2} =& \ \frac{1}{16 {\bar c}_s^2} \left(\frac{k_1k_2k_3}{2 K^3}\right)^{3-2\nu}  \left[\epsilon^2\, {\cal I}_{\epsilon^2 }(\alpha_2)\right]\,, \label{3ampI}
\end{align}

where

\begin{align} %\label{AcaseI}
{\cal I}_{\zeta \dot\zeta^2}(\alpha)\, =&\ \cos\frac{\alpha \pi}{2} \Gamma(1+\alpha) \left[(2+\alpha) \frac{1}{K}\sum_{i<j}k_i^2k_j^2  - (1+\alpha) \frac{1}{K^2} \sum_{i\neq j} k_i^2 k_j^3\right] \nonumber\;; \\[2.5mm] \nonumber
{\cal I}_{\zeta(\partial \zeta)^2}(\alpha) =& - \cos\frac{\alpha\pi}{2} \Gamma(1+\alpha) \left(\sum_i k_i^2\right)\left[\frac{K}{\alpha -1} + \frac{1}{K} \sum_{i<j}k_i k_j + \frac{1+\alpha}{K^2} k_1 k_2 k_3\right]\\[2.5mm]
= & \cos\frac{\alpha\pi}{2} \Gamma(1+\alpha)\left[\frac{1}{1-\alpha} \sum_j k_j^3 +\frac{4+2\alpha}{K}\sum_{i<j}k_i^2k_j^2 \nonumber
 - \frac{2+2 \alpha}{K^2} \sum_{i\neq j} k_i^2 k_j^3 
 \right.\\
 &  \nonumber \left. 
 + \frac{\alpha}{(1-\alpha)} \sum_{i\neq j} k_i k_j^2  - \alpha k_1 k_2 k_3 \right] \;;  \\[2.5mm]
{\cal I}_{\dot \zeta \partial \zeta \partial \chi}(\alpha) =& \cos\frac{\alpha \pi}{2}\Gamma(1+\alpha) \left[\sum_j k_j^3  + \frac{\alpha -1}{2}\sum_{i\neq j}k_i k_j^2 - 2 \frac{1+\alpha}{K^2}\sum_{i\neq j}k_i^2 k_j^3 - 2 \alpha k_1 k_2 k_3\right]\;; \nonumber \\[2.5mm]
{\cal I}_{\epsilon^2} (\alpha)= &  \cos\frac{\alpha \pi}{2} \Gamma(1+\alpha)(2+\alpha/2)\left[\sum_j k_j^3 - \sum_{i\neq j} k_i k_j^2 + 2  k_1 k_2 k_3\right]\,. \label{amplitudes}
%\label{3ampI}
\end{align}
In the above, ${\cal A}_{\epsilon ^2}$ accounts for the $\partial \zeta \partial \chi \partial^2 \chi$ and  $(\partial^2 \zeta) (\partial \chi)^2$ terms and the first term in ~\eqref{action3} has not been considered because it is identically null in the case of a DBI-type action.

One should note that, in contrast to the $c_s \to \infty$ limit relevant in the bimetric context, in the subluminal case ($c_s<1$), it is in fact the ``$c_s^{-2}$" terms that dominate.

\end{document}